\documentclass[aps,prl,superscriptaddress,twocolumn,nopacs,letter]{revtex4-1}
\usepackage{graphicx,bm,times}
\usepackage{amsmath}
\usepackage{amsfonts}
\usepackage{amssymb}
\usepackage{color}
\usepackage{hyperref}
\hypersetup{
	colorlinks = true,
	allcolors = {blue}
}
\usepackage[utf8]{inputenc}
\usepackage[T1]{fontenc}
\usepackage{graphicx}
\usepackage{amsmath,amssymb}
\usepackage{lineno}
\usepackage{float}
\usepackage{siunitx}

\begin{document}

\title{Bulk and surface electronic states in the dosed semimetallic HfTe$\boldsymbol{_2}$}

\author{Zakariae El Youbi}
\affiliation {Diamond Light Source, Harwell Campus, Didcot, OX11 0DE, United Kingdom}
\affiliation {Laboratoire de Physique des Matériaux et des Surfaces, CY Cergy Paris Université, 5 mail Gay-Lussac, 95031 Cergy-Pontoise, France}
\author{Sung Won Jung}
\affiliation {Diamond Light Source, Harwell Campus, Didcot, OX11 0DE, United Kingdom}
\author{Saumya Mukherjee}
\affiliation {Diamond Light Source, Harwell Campus, Didcot, OX11 0DE, United Kingdom}
\affiliation{Clarendon Laboratory, Department of Physics, University of Oxford, Parks Road, Oxford OX1 3PU, United Kingdom}
\author{Mauro Fanciulli}
\affiliation {Laboratoire de Physique des Matériaux et des Surfaces, CY Cergy Paris Université, 5 mail Gay-Lussac, 95031 Cergy-Pontoise, France}
\affiliation{Université Paris-Saclay, CEA, CNRS, LIDYL, 91191, Gif-sur-Yvette, France}
\author{Jakub Schusser}
\affiliation {Laboratoire de Physique des Matériaux et des Surfaces, CY Cergy Paris Université, 5 mail Gay-Lussac, 95031 Cergy-Pontoise, France}
\affiliation{New Technologies-Research Center, University of West Bohemia, 30614 Pilsen, Czech Republic}
\author{Olivier Heckmann}
\affiliation {Laboratoire de Physique des Matériaux et des Surfaces, CY Cergy Paris Université, 5 mail Gay-Lussac, 95031 Cergy-Pontoise, France}
\affiliation{Université Paris-Saclay, CEA, CNRS, LIDYL, 91191, Gif-sur-Yvette, France}
\author{Christine Richter}
\affiliation {Laboratoire de Physique des Matériaux et des Surfaces, CY Cergy Paris Université, 5 mail Gay-Lussac, 95031 Cergy-Pontoise, France}
\affiliation{Université Paris-Saclay, CEA, CNRS, LIDYL, 91191, Gif-sur-Yvette, France}
\author{J\'an Min\'ar}
\affiliation {New Technologies-Research Center, University of West Bohemia, 30614 Pilsen, Czech Republic}
\author{Karol Hricovini}
\affiliation {Laboratoire de Physique des Matériaux et des Surfaces, CY Cergy Paris Université, 5 mail Gay-Lussac, 95031 Cergy-Pontoise, France}
\affiliation{Université Paris-Saclay, CEA, CNRS, LIDYL, 91191, Gif-sur-Yvette, France}
\author{Matthew D. Watson}
\affiliation {Diamond Light Source, Harwell Campus, Didcot, OX11 0DE, United Kingdom}
\author{Cephise Cacho}
\email{cephise.cacho@diamond.ac.uk}
\affiliation {Diamond Light Source, Harwell Campus, Didcot, OX11 0DE, United Kingdom}

\date{\today}

\begin{abstract}
The dosing of layered materials with alkali metals has become a commonly used strategy in ARPES experiments. However, precisely what occurs under such conditions, both structurally and electronically, has remained a matter of debate. Here we perform a systematic study of 1T-HfTe$_2$, a prototypical semimetal of the transition metal dichalcogenide family. By utilizing photon energy-dependent angle-resolved photoemission spectroscopy (ARPES), we have investigated the electronic structure of this material as a function of Potassium (K) deposition. From the k$_z$ maps, we observe the appearance of 2D dispersive bands after electron dosing, with an increasing sharpness of the bands, consistent with the wavefunction confinement at the topmost layer. In our highest-dosing cases, a monolayer-like electronic structure emerges, presumably as a result of intercalation of the alkali metal. Here, by bringing the topmost valence band below $E_F$, we can directly measure a band overlap of $\sim$ 0.2 eV. However, 3D bulk-like states still contribute to the spectra even after considerable dosing. Our work provides a reference point for the increasingly popular studies of the alkali metal dosing of semimetals using ARPES.
\end{abstract}

\maketitle

\section{Introduction}

Alkali metal deposition on layered materials has turned into a commonly used approach in angle-resolved photoemission spectroscopy (ARPES) experiments, due to the possibility to stabilise and tune novel electronic states with properties that can significantly differ from those of the pristine material. It has been shown that dosing can suppress the charge density wave (CDW) phase in TiSe$_2$, and induce a metal-to-insulator transition alongside with a novel CDW phase in TaS$_2$ \cite{RossnagelNew_J._Phys2010}. In the semiconducting Black Phosphorus, the bandgap was tuned via K dosing, prompting the emergence of Dirac fermions due to the surface Stark effect \cite{KimScience2015, BaikNanoLett2015,KimPRL2017,JungNatmaterials2020}. This effect has been announced as a universal mechanism of band-gap engineering in 2D semiconductors \cite{KangNanoLett2017}. Besides, dimensionality reduction, whether upon alkali metal evaporation, hydrogenation or water vapor exposure has been widely reported \cite{EknapakulNanoLett2014,RileyNnano2015,ClarkPRB2019,NakataPRM2019,ChoPCCP2018,BeniaPRL2011}. 

It has been suggested that dosing semimetals could be qualitatively different to dosing semiconductors \cite{ClarkPRB2019}. In semiconductors, the most familiar scenario is ``band-bending", where a strong variation of the electrostatic potential near the surface results in the effective spatial confinement and emergence of quantum-well-like states \cite{BeniaPRL2011,BahramyNComms2012,RileyNnano2015,KangNanoLett2017}. Whereas in semimetals, the existence of both hole-like and electron-like free carriers makes the underlying mechanism complex to predict \cite{ClarkPRB2019}. On the other hand, in some cases the alkali metals are known to migrate into the van der Waals gaps of the sample, decoupling a top layer with a monolayer-like electronic structure \cite{EknapakulNanoLett2014}. 

1T-HfTe$_2$ is a transition-metal dichalcogenide (TMD) which has a semimetallic electronic structure, with both electronlike and holelike bands crossing the Fermi level \cite{HodulJ.Phys.Chem.Solids1985,KlipsteinJ._Phys1986,AokiSyntheticMaterials1995,AokiJ._Phys.Society.Japan1996,ReshakPhysicaB2005,Aminalragia2DMat2016,MangelsenPRB2017,NakataPRM2019}. Early studies of this material were focused on the fundamental properties such as the stoichiometry, lattice constants, transport properties and thermopower \cite{SmeggilJ._Solid.State.Chemistry1972,HodulPhysicaB1980,HodulJ.Phys.Chem.Solids1985,KlipsteinJ._Phys1986,AokiJ._Phys.Society.Japan1996,ReshakPhysicaB2005}. A recent ARPES study of thin films of HfTe$_2$ claimed that 1T-HfTe$_2$ might be classed as a topological Dirac semimetal \cite{Aminalragia2DMat2016}, though the bands observed in measurements of single crystals did not support this \cite{NakataPRM2019}. Still, 1T-HfTe$_2$ shows a notably large and non-saturating magnetoresistance, resulting from the carrier compensation \cite{MangelsenPRB2017}. Its electronic structure is closely analagous to that of TiSe$_2$, which exhibits a CDW and can host superconductivity upon electron doping, though no such phase transition have been detected in HfTe$_2$ to date \cite{KlipsteinJ._Phys1986,AokiJ._Phys.Society.Japan1996,MangelsenPRB2017}.

\begin{figure*}
	\centering
	\includegraphics[width=1\textwidth]{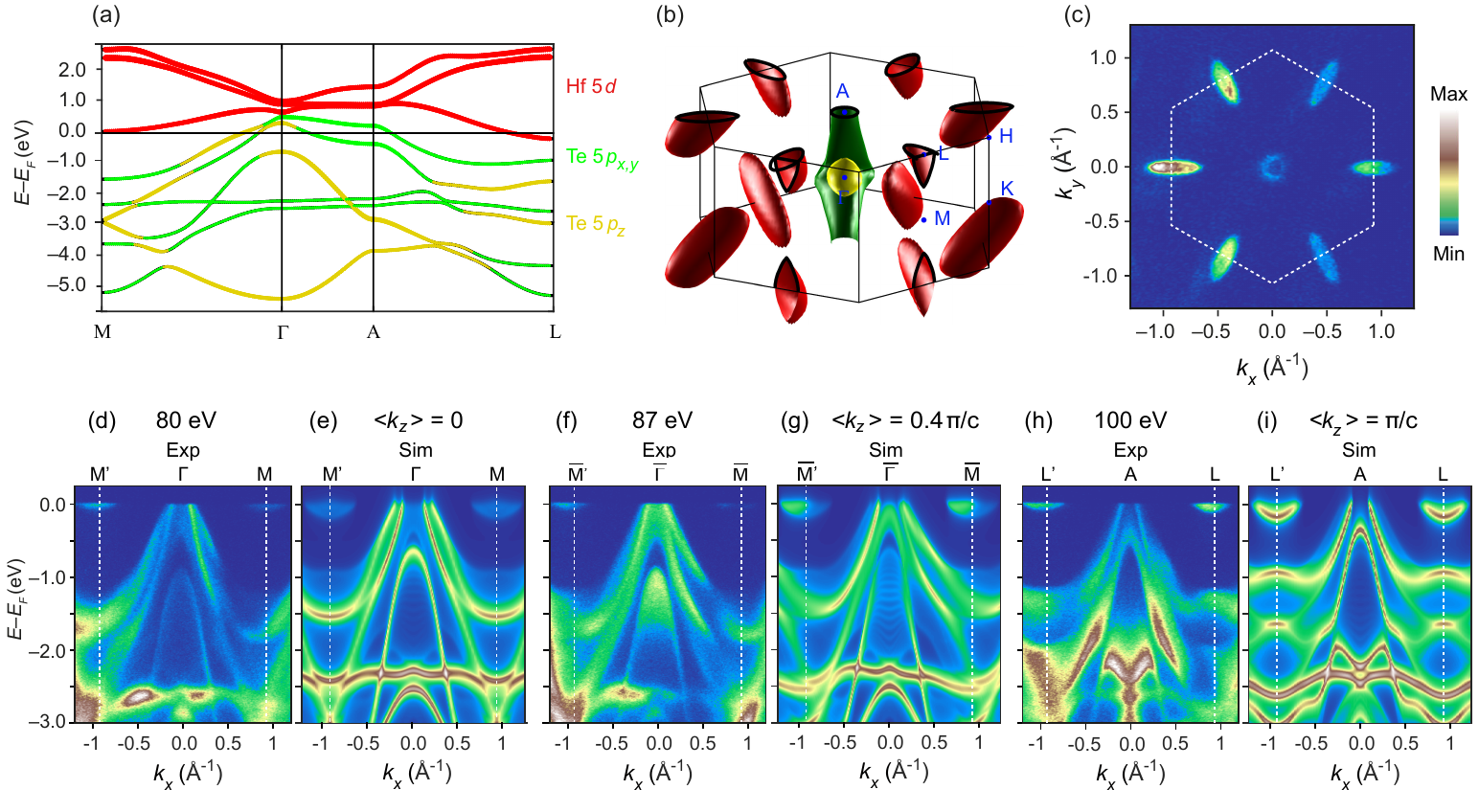}
	\caption{{Electronic structure of 1T-HfTe$_2$.} (a)  DFT calculations along M-$\Gamma$-A-L direction with orbital character projection of the valence and conduction bands. (b) 3D Fermi surface of HfTe$_2$ calculated with a DFT code using the mBJ functional. (c) Constant energy map at E$_F$ (Fermi surface), measured at a photon energy $h\nu=100$~eV to probe the A-H-L plane. The threefold in-plane intensity distribution of the electron pockets located at the L points reflect the trigonal symmetry of the crystal structure. (d,f,h) Valence band dispersion along $\Gamma$-M ($h\nu=80$~eV) (d), an arbitrary k$_z$ between $\Gamma$ and A point ($h\nu=87$~eV) (f) and A-L direction ($h\nu=100$~eV) (h). (e,g,i) Simulations based on DFT calculations, averaging over a substantial fraction of the Brillouin zone width in the $k_z$ direction, as further described in the main text. Panels (e),(g) and (i) correspond to an average $k_z$ value of \emph{$k_z$}=$~$0.0$\pi$/\emph{c}, \emph{$k_z$}=$~$0.4$\pi$/\emph{c}, and \emph{k$_z$}=$~\pi$/\emph{c} respectively. }
	\label{fig1}
\end{figure*}

In this paper, we combine high-resolution ARPES measurements and Density Functional Theory (DFT) calculations, which demonstrate the 3D character of the semimetallic ground state of HfTe$_2$. We then create 2D surface-confined states by sequentially dosing K metal ions on the surface. We observe an increasing sharpness of the bands with dosing level, since these states are not affected by the \emph{$k_z$} broadening effect. Our photon energy-dependent ARPES confirms the 2D nature of the surface-confined states after K evaporation, but we additionally observe a remnant bulk-like contribution. Interestingly, this weak \emph{k$_z$}-dispersive spectral weight manifests with a slightly different photon-energy dependence to the spectra of pristine sample. Our measurements establish HfTe$_2$ as a semimetal with a narrow band overlap (negative bandgap) of ~0.2 eV, and show how the bandstructure also supports a 3D Dirac point approximately 3 eV below $E_F$.

\section{Methods}

ARPES measurements were performed at the I05 beam line of Diamond Light Source \cite{HoeshRev.Sc.Instrum2017}. The photon energy was varied between 80 and 120 eV, and the light was linearly polarized in the horizontal plane (LH, or $p$ polarization). High quality HfTe$_2$ single crystals were obtained from HQ Graphene. Due to their air-sensitivity, the samples were prepared in a dry argon glove box. Samples were cleaved \emph{in situ} and measured at a temperature of $T=10$~K. Potassium K was deposited on the cleaved surface at a low rate using a well-outgassed SAES K getter source operated at current of 6 A. The energy resolution was typically 10 meV. 

DFT calculations were performed within the Wien2k package \cite{Blahawien2k2001}, accounting for spin-orbit coupling. The modified Becke-Johnson (mBJ) functional \cite{KollerPRB2012} was used (see text for discussion). The used lattice parameters are $a=b=3.911 \si{\angstrom}$, $c=6.649 \si{\angstrom}$ and $z_{Te}=0.266$ \cite{MangelsenPRB2017}, and the R$_{MT}K_{max}$ parameter was equal to 7.0.

\section{Results}

HfS$_2$ and HfSe$_2$ are well-established as layered semiconducting materials \cite{kanazawaSc.reports2016,ZhaoPhys.Status.Soli.2017}. In the case of HfTe$_2$, however, the more extensive Te 5\emph{p} orbitals have a larger overlap with the Hf $5d$ orbitals, and results in a semimetallic ground state. Thus, the electronic structure of 1T-HfTe$_2$ has more in common with isostructural and isovalent 1T-TiTe$_2$, a well-known semimetal \cite{ClaessenPRB1996,RossnagelPRB2001,ReshakPRB2003}. Our DFT calculations with the orbital character projection (Fig.~\ref{fig1}(a)) gives an overview of the electronic structure. We have a manifold of 6 bands deriving from Te 5\emph{p} orbitals, which are mostly occupied but reach up to the Fermi level around the Brillouin zone center ($\Gamma$ and A points), forming hole pockets. Conversely, the bands deriving from Hf 5\emph{d} orbitals are mostly unoccupied, but dip below the Fermi level to form electron pockets around the L points. Thus HfTe$_2$ is a compensated semimetal, though the hole and electron pockets have very different characters.

Although HfTe$_2$ is a layered material, the out-of-plane dispersion is crucial to understand the full three-dimensional electronic structure. In the calculation in Fig.~\ref{fig1}(a) at the A point, two hole-like bands are present near $E_F$. These derive from Te 5\emph{p$_{x,y}$} orbitals only, but the separation of the bands due to the spin-orbit coupling is so large ($\sim$ 0.65~eV) that the lower branch is shifted fully into the occupied states, leaving a single hole pocket here. At the $\Gamma$ point, three bands are present as the $5p_z$ orbital is also relevant, but again there is a strong spin-orbit interaction which mixes the orbital character and separates the bands; the result is a pair of hole bands crossing $E_F$ and a lower hole band which remains fully occupied. Thus, as illustrated in the 3D Fermi surface in Fig.~\ref{fig1}(b), only one hole band crosses $E_F$ at the A point (green band), but two Fermi pockets exist around $\Gamma$ (green and yellow bands).

Similar to the widely-known TiSe$_2$, the electron pockets are centered at the L points in HfTe$_2$. Whether or not the Fermi surface is closed (i.e. a 3D pocket, as in TiSe$_2$ and TiTe$_2$ \cite{WatsonPRL2019,RossnagelPRB2001}) or open along \emph{k$_z$} (i.e. forming a warped 2D cylinder) is a subtle question, but important for the understanding of transport data \cite{MangelsenPRB2017,MangelsenInorg.Chem2019}. In DFT calculations using the Generalized Gradient Approximation (GGA) functional the pockets are open along \emph{k$_z$}, but GGA calculations also give an unrealistically large overlap of the Te 5\emph{p} and Hf 5\emph{d} states. Using the mBJ functional, however, gives a much more accurate low-energy band structure, and yields a closed electron pocket around the L points; or in other words, the lowest Hf 5\emph{d} state at the M point in Fig.~\ref{fig1}(a) is slightly above $E_F$. 

How this 3D electron pocket appears in ARPES measurements requires some detailed discussion, because the \emph{k$_z$} dispersion of the initial state is clearly as significant as the \emph{$k_x,k_y$} directions; however, \emph{k$_z$} is not strictly conserved in the photoemission process \cite{Damascelli_2004}. In general, the existence of multiple possible final states, and their non-parabolic dispersions, makes the full treatment of photoemission a complex problem. A commonly-used approximation is the so-called free-electron final state model, which assumes a simplistic dispersion for the final state that allows a mapping between the photoelectron kinetic energy and the $k_z$ value of the initial state. Empirically, we find this approximation to be largely successful for HfTe$_2$, at least for the data here obtained with photon energies on the order of 100~eV. However, our measurements at a given photon energy appear to probe a distribution in $k_z$, rather than a sharply-defined value. This ``uncertainty" in $k_z$ is generically expected due to the damping of the final state in the $z$ direction \cite{Strocov2003, StrocovPRB2006}, normal to the sample surface. Such $k_z$ broadening is evident from the constant energy map in Fig.~\ref{fig1}(c), measured at a photon energy $h\nu=100$~eV to probe the A-H-L plane. Here, the electron pockets are ``filled-in", with intensity localised within the bounds of the electron pocket, but we do not observe sharp ellipsoidal countours that might be expected from slicing the 3D Fermi surface in the $k_z=\pi/c$ plane. The effect of $k_z$ uncertainty is more pronounced for the electron pockets as they have a slanted 3D shape. Whereas the hole pocket is centered on an axis of rotation which forbids any such slanting, and thus appears relatively sharp at the A point. 

The measured high-symmetry dispersions are also influenced by the uncertainty in \emph{k$_z$}. Therefore, we do not simply compare the measurements with the calculated dispersions from DFT, which assume a particular \emph{k$_z$} value. Rather, we simulate the effect of \emph{k$_z$} broadening by averaging over a substantial fraction of the Brillouin zone. We first construct a spectral function for each calculated $k_z$ by adding a finite scattering rate of 50 meV to the calculated in-plane band dispersions, and then perform a summation over $k_z$ in which the weight of each $k_z$ slice is modelled by a Lorentzian distribution with FWHM of $0.2 \times (\frac{2\pi}{c})$. Since the 80 eV data in Fig.~\ref{fig1}(d) corresponds to a $\Gamma$ point, according to the free-electron final state approximation, the simulation in Fig.~\ref{fig1}(e) is centered around \emph{k$_z$} =$~$0. Matrix elements are not included in the simulation, and the variation of $k_z$ with in-plane momentum is not accounted for. Moreover, the calculated band energies are typically $\sim$100 meV offset compared to the experiments, and correspondingly both the hole and electron pockets are larger in the calculation than in the experiment. However, there is excellent correspondence in understanding where features are sharp in the data (e.g. the inner hole pocket) or rather broader, due to the $k_z$ averaging effect \cite{MitsuhashiPRB2016}. Some weak spectral weight of the electron pockets appear at the M points, however, this should be understood strictly as a result of smearing of spectral weight along $k_z$, and not as quasiparticle bands existing at M. At the L points, in Fig.~\ref{fig1}(h), the electron pockets are brighter and better defined, though somewhat smaller than in the simulation based on DFT projection in Fig.~\ref{fig1}(i), due to the limitations of the accuracy of the functional. 

The merit of this approach is seen best for data measured away from any high-symmetry \emph{k$_z$} value, such as in Fig.~\ref{fig1}(f). Here, the data show a substantial asymmetry, as well as a large degree of broadening. However, these complexities are well-captured by the simulation in Fig.~\ref{fig1}(g), where the \emph{k$_z$} distribution is centered at 0.4$\pi/c$. The asymmetry in the spectrum arises because of the $\bar{3}$ symmetry; away from the high-symmetry $k_z$=(0, $\pm{}\pi{}/c$) planes, band dispersions towards the $\bar{M}$ and $\bar{M'}$ directions are not equivalent. The simulation also shows that the spectral weight of the electron pocket is centered away from the $\bar{M}$ point, which is due to the slanting of the electron pockets. The spectral weight is shifted inside the Brillouin zone near the $\bar{M}$ point and outside of the first Brillouin zone at the $\bar{M'}$ point. Therefore, in the general case we expect threefold and not sixfold symmetry in measured ``Fermi maps", which is already present to some degree in Fig.~\ref{fig1}(c) but is pronounced in measurements at other photon energies \cite{TangPRB2019}. Overall, then, the agreement with our simulation based on DFT calculations confirms that the observed features are projected bulk bands resulting from the k$_z$ uncertainties during the photoemission process. 

\begin{figure}
    \centering
	\includegraphics[width=0.95\columnwidth]{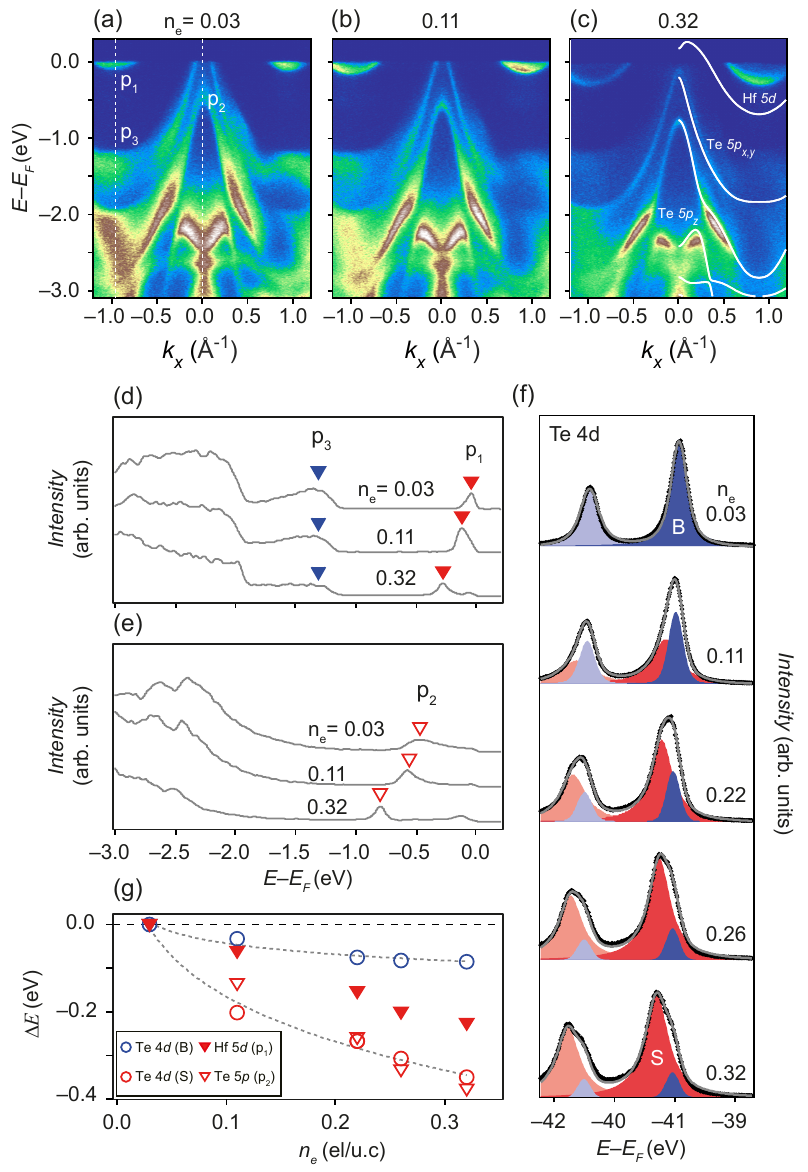}
    \caption{{Dosing sequence of HfTe$_2$ at $T=10$~K, measured at a photon energy of $h\nu=100$~eV.} (a) Valence band dispersion along A-L direction of the pristine surface, (b) at a carrier density of n$_e$=0.11  and (c) at n$_e$=0.32 el/u.c., together with HfTe$_2$ monolayer DFT calculations (white solid lines), shifted downwards by 650 meV. Due to the overestimated band overlap in DFT, the energy of the electron pocket is lower in the calculation than the data, however the agreement on the valence band structure is more important here. (d) Energy Distribution Curves (EDCs) showing the evolution as a function of the charge carrier density. The labelled features correspond to the electron pockets at the L point (p$_1$), and the remnant bulk-like bands (p$_3$). (e) Equivalent EDCs at A, highlighting a peak from the inner valence band (p$_2$). (f) Core level spectroscopy of Te 4\emph{d} as a function of charge carrier density. The spectra were fitted using two pairs of Voigt functions, depicted as blue (bulk) and red (surface) shaded areas. The SOC peak separation and areal ratio of Te \emph{4d$_{3/2}$} to Te \emph{4d$_{5/2}$}, were fixed to the values obtained from the pristine surface (n$_e$=$0.03$). (g) Comparison of the energy shift of shallow core levels of Te 4\emph{d} (blue and red circles), the inner Te 5\emph{p}-derived state of the valence band (p$_2$, filled red triangles) and the Hf 5\emph{d}-derived electron pockets (p$_1$, empty red triangles). The data reveals a similar trend of the p$_2$ valence band state (Te $5p$) and Te 4\emph{d} shallow core levels.}
    \label{fig2}
\end{figure}

We now turn to the question of how a 2D electronic structure can emerge upon K dosing of the surface of HfTe$_2$. At first glance, the ARPES data in Fig.~\ref{fig2}(a-c) might seem like a simple rigid shift of the band structure, where the electron pockets start to become substantially occupied and the hole pockets eventually shift fully into the occupied states, reaching an effective carrier density of $n_e$=0.32 el/u.c in Fig.~\ref{fig2}(c). However, there are a number of important effects beyond the rigid shift model. First, the bands in the heavily-dosed spectra (Fig.~\ref{fig2}(c)) appear sharper than in the pristine bulk case (Fig.~\ref{fig2}(a)). The increased sharpness is somewhat counter-intuitive, since in many cases surface dosing (and doping more generally) goes hand-in-hand with increased disorder, thereby leading to broader, not sharper, linewidths \cite{KangNanoLett2017}. We must consider, however, that for the pristine surface, the $k_z$-broadening effect is the dominant source of linewidth broadening for most of the bands. The increasing sharpness is therefore the first experimental signature that the states that emerge after dosing have 2D character, with wavefunctions confined to the topmost layer of HfTe$_2$, thus suppressing the $k_z$-broadening. Similar phenomenenology has been observed in other TMDCs with more 3D characteristics in their bulk electronic structure \cite{ClarkPRB2019}. The sharpness also confirms that the K coverage must be homogeneous, at least on the scale of the beam spot size.


A recurrent question in the study of alkali-metal dosed surfaces of 2D materials is whether the dosed atoms remain on the surface and set up a vertical electric field, or intercalate into the sample \cite{RossnagelNew_J._Phys2010,KimScience2015,BaikNanoLett2015,KimPRL2017,KangNanoLett2017,EknapakulNanoLett2014,RileyNnano2015,ClarkPRB2019,NakataPRM2019,ChoPCCP2018,BeniaPRL2011,BahramyNComms2012,EknapakulPRB2016,AlidoustNcomms2014, EknapakulPRB2018,Schroder2Dmaterials2016,EhlenPRB2018}. A number of recent studies on TMDCs and similar materials have claimed that the alkali metal atoms remain on the surface \cite{RileyNnano2015,KimScience2015,BaikNanoLett2015, KimPRL2017,KangNanoLett2017,EhlenPRB2018,ClarkPRB2019,JungNatmaterials2020}, in the case that the sample remains at low temperatures throughout. In this scenario, the alkali metal ions create a charge accumulation at the surface, causing a band-bending potential that affects primarily the top-most layer of the TMDC. The new bands that emerge thus appear electron-doped compared with the bulk, and their wavefunctions are localised close to the surface \cite{BahramyNComms2012,ClarkPRB2019}. Without detailed calculations, we cannot specifically predict what band dispersions would be expected in our case, however it would be reasonable to imagine that, broadly speaking, the conduction band will become increasingly filled, while the observed valence bands would develop more 2D character.

The alternative scenario is intercalation, in which the K atoms migrate into the van der Waals gap between layers, leading to an increased van der Waals gap that causes decoupling of the topmost layer both structurally and electronically. This scenario is known to occur almost immediately at room temperature \cite{RossnagelNew_J._Phys2010} but is also often applied to dosing studies of TMDCs even when the sample is held at low temperatures \cite{EknapakulNanoLett2014,EknapakulPRB2018,NakataPRM2019}. This topmost layer also becomes doped due to the donated electrons from the K$^+$ ions.

The two scenarios lead to somewhat similar effects electronically, since in both cases electron-doped 2D surface-confined states emerge (i.e. wavefunctions localised in the topmost layer of HfTe$_2$). The vertical field scenario is a plausible scenario for our low-dosing data in Fig.~\ref{fig2}(b).
Indeed, the data in Fig.~\ref{fig2}(b) largely resembles the bulk band structure, but with an overall energy shift, which could be accounted for by a band-bending potential at the surface. Whereas after further dosing the appearance of the data changes qualitatively (the first point beyond $n_e$=0.11 el/u.c in our dosing series), especially in the structure and intensity distribution of the bands around -2.5 eV. Moreover, we find very good agreement overall between the shape of the valence bands in Fig.~\ref{fig2}(c) and the DFT calculations for an isolated monolayer \footnote{For the monolayer calculation, we fix the $c$ parameter of the mBJ functional using the bulk value, rather than calculating it during the self-consistent cycle.}. This is evidence in favour of the intercalation scenario, as it implies that the topmost layer is structurally decoupled, suppresses out-of-plane hopping, which particularly affects the $p_z$ orbital, leading to the emergence of monolayer-like dispersions of the valence bands. Thus at least for our highest-dosing samples, we are in agreement with Ref.~\cite{NakataPRM2019} that the intercalation scenario is most likely.

For a more detailed analysis we extract energy distribution curves (EDCs) from the L and A points and track their evolution with K dosing. The feature $p_1$ in Fig.~\ref{fig2}(d) shows a shift of $\sim$ 0.22 eV to lower energy, reflecting the filling of the Hf 5\emph{d}-derived conduction band at L. The inner Te 5\emph{p}-derived hole pocket in Fig.~\ref{fig2}(e), $p_2$, shows qualitatively a similar shift but with a slightly larger magnitude of $\sim$ 0.37 eV. The different behavior of these bands as a function of K deposition at the surface may be related to their different orbital character. Assuming that the shift of $p_2$ is representative for the hole bands overall, the different magnitude of the shift implies that the band overlap decreases somewhat with dosing. This indicates that a rigid shift model is not applicable in this material. The lineshape of $p_2$ in Fig.~\ref{fig2}(e) also becomes noticeably sharper, as the surface confinement takes effect and the state loses any $k_z$-dependence. However, other features such as $p_3$ simply lose intensity (though still weakly contributing even at high dosing) and show low or negligible energy shifts with increasing dosing in (Fig.~\ref{fig2}(d)). These correspond to remnant intensity from bulk states, which we will focus on later.

In addition to the shifts of the Te 5\emph{p}-derived valence bands, the Te 4\emph{d} shallow core levels also show a significant change with K dosing (Fig.~\ref{fig2}(f)). Consistent with a single environment for the Te atoms, these appear as a sharp doublet in the pristine case. As we dose the surface with K, however, fitting the data using two Voigt function pairs becomes necessary, reflecting the separate chemical shifts in the surface layer and the bulk. The intensity of the surface contribution (red) relative to the bulk-like (blue) signal increases with increasing dosing, concomitant with their energetic separation increasing also. Fig.~\ref{fig2}(g) shows that the energetic shift of the Te 4\emph{d} state has approximately the same tendency with the shifts of the valence bands. 
Thus overall, as has been pointed out by \cite{NakataPRM2019}, this electronic structure may be understood as being close to a doped monolayer of 1T-HfTe$_2$, with the Te 4\emph{d} levels experiencing a similar shift of the chemical potential to the valence bands, but the Hf states showing a smaller shift.


\begin{figure*}
	\centering
	\includegraphics[width=1\textwidth]{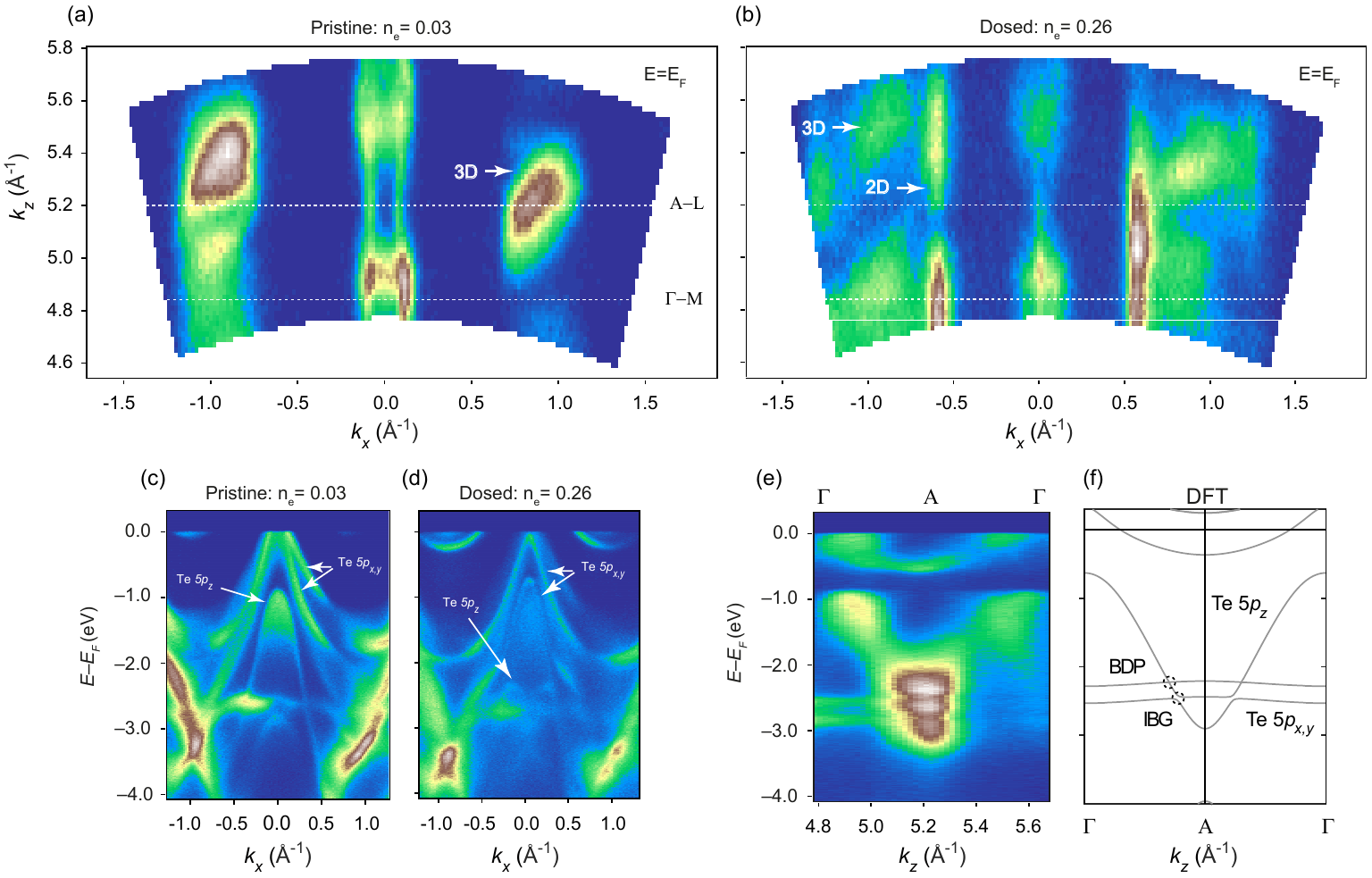}
	\caption{{Photon energy-dependent ARPES.} (a)-(b) \emph{k$_z$} maps, processed from photon-energy dependent data at E$_F$ before and after Potassium dosing on HfTe$_2$ surface along M-$\Gamma$-M’ [L-A-L’] direction, providing clear evidence of the emergence of surface-confined states and the remnant bulk-like contribution due to the non-negligible chance of the deep photoelectrons to escape. Photon energies into \emph{k$_z$} conversion was done by employing the standard free-electron final state approximation: $\emph{k$_z$} = \frac{1}{\hbar} \sqrt{2m_e(V_0 + E_k \cos^2\theta)}$, where $\theta$ is the in-plane emission angle, $\emph{E$_k$}$ is the photo-electron kinetic energy and $\emph{V$_0$}$ is the inner potential. We use an inner potential of 11$~$eV. (c)-(d) Valence band dispersion at a photon energy $h\nu=87$~eV of the pristine and dosed surface, showing unambiguously the asymmetry disappearance and sharpness of the bands due the monolayer-like electronic structure after dosing. The observed features inside the electron pockets and the center are a clear evidence of the remnant bulk states of deeper photoelectrons in the material. (e)-(f) Band dispersion along the out-of-plane \emph{k$_z$} direction at the Brillouin zone centre [$\Gamma$-A direction] of the pristine surface, highlighting the expected crossings of the Te \emph{p$_z$} orbital and Te \emph{p$_{xy}$} orbital-derived states, giving rise to a bulk Dirac point (BDP) (upper crossing) and an inverted bandgap (IBG) (lower crossing), as labeled in our DFT calculations.}
	\label{fig3}
\end{figure*}

Figure \ref{fig3}(a,b) shows photon-energy-dependent ARPES measurements before and after K deposition on HfTe$_2$, mapped into \emph{k$_z$} using the free electron final state approximation. The \emph{k$_z$} map of the pristine surface shows a pair of Te 5\emph{p$_{xy}$} with the contribution 5\emph{p$_z$}-orbital derived state at the $\Gamma$ point, and electron pockets centered at each L point (Fig.~\ref{fig3}(a)). The asymmetry of these pockets reflects, however, the trigonal symmetry of the crystal and the matrix element effects during the photoemission process. Nevertheless, all the states forming the band structure of the pristine surface have a 3D character, as discussed in details above.

Upon alkali metal deposition, we observe the emergence of states without $k_z$ variation, which confirms that the electron pockets in the surface-dosed system are strictly 2D surface-confined states. In addition to these 2D states, there exists a noticeable contribution of 3D bulk states as shown in Fig.~\ref{fig3}(b). These resemble the 3D electronic structure of the pristine surface, though with some subtle shifts. We interpret these weaker features as photoelectrons being excited deeper in the material and have a non-negligible chance to escape, analogous to the remnant bulk-like contribution to the core level spectra in Fig.~\ref{fig2}(e). Interestingly, the spectral weight of these remnant 3D states appears to be shifted along the $k_z$ axis (using the same value of the inner potential for the $k_z$-mapping). In other words, a modified inner potential may be applicable for these photoelectrons. Within the three-step model, the inner potential includes a term from the material work function \cite{DamascelliRMP2003,Damascelli_2004,Hufner2013}, and this is known to be strongly dependent on K dosing \cite{JaouenArxiv2019}. We speculate therefore that even though the initial states of this bulk-like contribution are essentially unaffected by the dosing, the photoelectrons experience a different surface potential, and thus come out with a different effective $k_z$.

In addition to the increased sharpness of the electronic structure, due to their 2D character after dosing, the asymmetry present in bulk spectra disappears in the surface-confined states (Fig.~\ref{fig3}(c,d)). This occurs because in the 2D limit, the band structure is expected to be sixfold symmetric. However, the weaker, asymmetric, and broader, spectral features come from a remnant contribution of the bulk states. In particular, our understanding of the spectral weight inside the electron pockets in Fig.~\ref{fig3}(d) comes from such bulk-like states, as evidenced from their 3D character in the \emph{k$_z$} map after dosing (Fig.~\ref{fig3}(b)). Similar bulk-like features are often observed even after a substantial dosing is applied \cite{WenNat.Comms2016}. 

One of the interesting features of the 3D electronic structure is that it supports the existence of the bulk Dirac fermions as well as an inverted bandgap. These topologically-protected features arise from within the manifold of Te \emph{p}-orbitals \cite{BahramyNat.Materials2018}. In this layered material, with a trigonal crystal field,  \emph{p}-orbital states form dispersive bands due to bonding, crystal-field splitting and spin-orbit coupling. In the out-of-plane \emph{k$_z$} direction, a band with Te \emph{p$_z$} orbital character behaves very differently to the in-plane \emph{p$_{xy}$}-derived states. This gives a general expectation of a strongly \emph{k$_z$} dispersion of the out-of-plane Te \emph{p$_z$} orbital and a dispersionless pair of in-plane \emph{p$_{xy}$}-derived states, thus naturally causing a set of \emph{k$_z$}-dependent crossings \cite{BahramyNat.Materials2018,ClarkPRL2018,ClarkElec.Structure2019,GhoshPRB2019,MukherjeeArxiv2019}. 

Indeed, our valence band dispersion between $\Gamma$ (\textbf{k} =$~$(0,0,0)) and A (\textbf{k} =$~$(0,0,$\pi$/\emph{c})) high-symmetry points of the Brillouin zone shows the \emph{p}-orbital-derived manifold of states along the out-of-plane \emph{k$_z$} direction (fig.~\ref{fig3}(e,f)). The crossing of the \emph{p$_z$} orbital-derived state with the upper \emph{p$_{xy}$} derived state forms a symmetry-protected 3D Dirac point, due to their belonging to different irreducible representations. However, \emph{p$_z$} and the lower \emph{p$_{xy}$} derived state share the same symmetry character and angular momentum but opposite parity, their hybridization thus opens an inverted bandgap which is expected to host topological surface states \cite{BahramyNat.Materials2018,ClarkElec.Structure2019}. The location of these crossings at deeper binding energies $\emph{E-E$_F$}$ $\sim$ 3$~$eV, and the small size of the inverted bandgap makes the resolution of such features challenging due to the broadening of the bands.

One of the advantages of dosing this semimetallic system is that, by bringing the uppermost hole band below $E_F$, the band overlap can be directly measured. From Fig.~\ref{fig2}(c), we estimate this to be $\sim$ 0.2 eV. We caution that this is a value for the surface-dosed system, and thus the bulk value may vary slightly, but we estimate it to be in the range 0.2-0.3 eV. A small band overlap is an interesting playground for novel electronic states. For instance, one could imagine substituting Se onto the Te site in order to continuously tune the band gap/overlap through zero, and it would be interesting to see if any electronic instabilities occur, such as in the well-known isovalent compound TiSe$_2$ \cite{WatsonPRL2019}.

\section{Conclusions}

To conclude, our findings show how the electronic structure of the HfTe$_2$ has undergone significant changes upon alkali metal deposition at the surface. The measured ARPES spectra revealed a 3D semimetallic ground state in the pristine sample. The broadening of the bands and asymmetries in the spectral weight were ascribed to the \emph{k$_z$} uncertainties in the photoemission process and the crystal structure, respectively. This then evolves into 2D surface-confined states, with correspondingly sharper linewidths and more symmetric intensity. After substantial doping, a monolayer-like electronic structure was revealed, indicating intercalation of the alkali metal and decoupling of the topmost layer of HfTe$_2$. However, we also found evidence of bulk-like features, arising from photoelectrons from deeper in the material, even after substantial K dosing. Furthermore, we show that HfTe$_2$ hosts topological features such as a BDP and an IBG, a further example of the generic mechanism described for other TMDCs in the same space group \cite{BahramyNat.Materials2018}.  Our methodical study of this prototypical layered semimetal gives evidence of the complexity that can occur upon alkali-metal dosing, showing qualitatively different behaviour to recent studies in another semimetal, PtSe$_2$ \cite{ClarkPRB2019}. Finally, we suggest that tuning the electronic structure of the narrow-band-overlap material HfTe$_2$ could lead to the discovery of novel electronic ground states. 

\section{Acknowledgments}

We thank Timur K. Kim and Benjamin Parrett for useful discussions. We thank Diamond Light Source for access to Beamline I05 (Proposal No. NT24921-1) that contributed to the results presented here. Z. El Youbi acknowledges Diamond Light Source and the university of Cergy-Pontoise for PhD studentship support. S. Mukherjee acknowledges financial support from the European Unions Horizon 2020 research and innovation programme under the Marie Skodowska-Curie Grant Agreement (GA) No. 665593 awarded to the Science and Technology Facilities Council. M. F. acknowledges support by the Swiss National Science Foundation, project no. P2ELP2-181877. J. S. and J. M. would like to thank CEDAMNF project financed by the Ministry of Education, Youth and Sports of Czech Repuplic, Project No. CZ.02.1.01/0.0/0.0/15.003/0000358. 

%


%

\end{document}